\documentclass[lineno]{biometrika}

\nolinenumbers

\usepackage{amsmath}
\DeclareMathOperator*{\argmax}{arg\,max}

\usepackage{amsfonts}
\usepackage{hyperref}
\usepackage{caption}
\usepackage{times}
\usepackage{bm}
\usepackage{natbib}
\usepackage{xcolor}

\usepackage{newtxtext}
\usepackage[subscriptcorrection]{newtxmath}
\usepackage{xr}

\graphicspath{{./art/}}

\usepackage[plain,noend]{algorithm2e}

\makeatletter
\renewcommand{\algocf@captiontext}[2]{#1\algocf@typo. \AlCapFnt{}#2} 
\def\@algocf@capt@plain{top}
\renewcommand{\algocf@makecaption}[2]{%
  \addtolength{\hsize}{\algomargin}%
  \sbox\@tempboxa{\algocf@captiontext{#1}{#2}}%
  \ifdim\wd\@tempboxa >\hsize
    \hskip .5\algomargin%
    \parbox[t]{\hsize}{\algocf@captiontext{#1}{#2}}
  \else%
    \global\@minipagefalse%
    \hbox to\hsize{\box\@tempboxa}
  \fi%
  \addtolength{\hsize}{-\algomargin}%
}
\makeatother

\begin{document}

\jname{Submitted to Biometrika}

\markboth{JIAZHEN Xu et~al.}{Change point detection for random objects with periodic behaviour}

\title{Change point detection for random objects with periodic behaviour}

\author{JIAZHEN XU}
\affil{Research School of Finance, Actuarial Studies and Statistics, Australian National University, Canberra ACT 2600, Australia
\email{jiazhen.xu@anu.edu.au}}

\author{ANDREW T.A. WOOD}
\affil{Research School of Finance, Actuarial Studies and Statistics, Australian National University, Canberra ACT 2600, Australia \email{andrew.wood@anu.edu.au}}

\author{TAO ZOU}
\affil{Research School of Finance, Actuarial Studies and Statistics, Australian National University, Canberra ACT 2600, Australia
\email{tao.zou@anu.edu.au}}

\maketitle

\begin{abstract}
Time-varying random objects have been increasingly encountered in modern data analysis.  Moreover, in a substantial number of these applications, periodic behaviour of the random objects has been observed. We develop a novel procedure to identify and localize abrupt changes in the distribution of non-Euclidean random objects with periodic behaviour. The proposed procedure is flexible and broadly applicable, accommodating a variety of suitable change point detectors for random objects. We further construct a specific detector used in the proposed procedure which is nonparametric and effectively captures the entire distribution of these random objects. The theoretical results cover the limiting distribution of the detector under the null hypothesis of no change point, the power of the test in the presence of change points under local alternatives and the consistency in estimating the number and locations of change points, whether dealing with a single change point or multiple ones. We demonstrate that the most competitive method currently in the literature for change point detection in random objects is degraded by periodic behaviour, as periodicity leads to blurring of the changes that this procedure aims to discover. Through comprehensive simulation studies, we demonstrate the superior power and accuracy of our approach in both detecting change points and pinpointing their locations. Our main application is to weighted networks, represented through graph Laplacians. The proposed method delivers highly interpretable results, as evidenced by the identification of meaningful change points in the New York City Citi Bike sharing system that align with significant historical events.
\end{abstract}

\begin{keywords}
Metric distribution function; Multiple change point detection; Non-Euclidean data; Nonparametric test; Transportation networks; $U$-statistics.
\end{keywords}

\section{Introduction}

Change point analysis has a rich history with applications in diverse areas such as quality control (\citealt{shewhart1986statistical,lai1995sequential,hawkins2003changepoint}), biology (\citealt{olshen2004circular,erdman2008fast,fearnhead2019changepoint,jiang2023time}), stock market analysis (\citealt{andreou2002detecting,bai2003computation,fryzlewicz2014wild}), climate and environmental studies (\citealt{rodionov2004sequential,reeves2007review,beaulieu2018distinguishing}) and social sciences (\citealt{kossinets2006empirical,wang2017fast}), to name but a few. The core aim of change point analysis is the identification and precise localization of abrupt changes in the distribution of observations that are ordered in some way, usually by time.

Traditionally, change point detection techniques have been developed for Euclidean data; see e.g., \cite{basseville1993detection}, \cite{aminikhanghahi2017survey} and \cite{truong2020selective}.  However, data that do not reside in Euclidean spaces are becoming increasingly common. Such non-Euclidean observations, often referred to as random objects (\citealt{marron2021object}), are neither scalars nor vectors, but instead take values in a general metric space which will often have non-Euclidean structure. Important examples include transportation networks, compositional data, functional Magnetic Resonance Imaging, phylogenetic trees; see e.g., \cite{worsley2002general}, \cite{holmes2003statistics}, \cite{kolar2010estimating} and \cite{scealy2023score}. The absence of vector operations such as addition, scalar multiplication, or inner products in such spaces poses significant challenges in the statistical analysis. 

\begin{figure}[htbp!]
\centering
\includegraphics[scale=0.42]{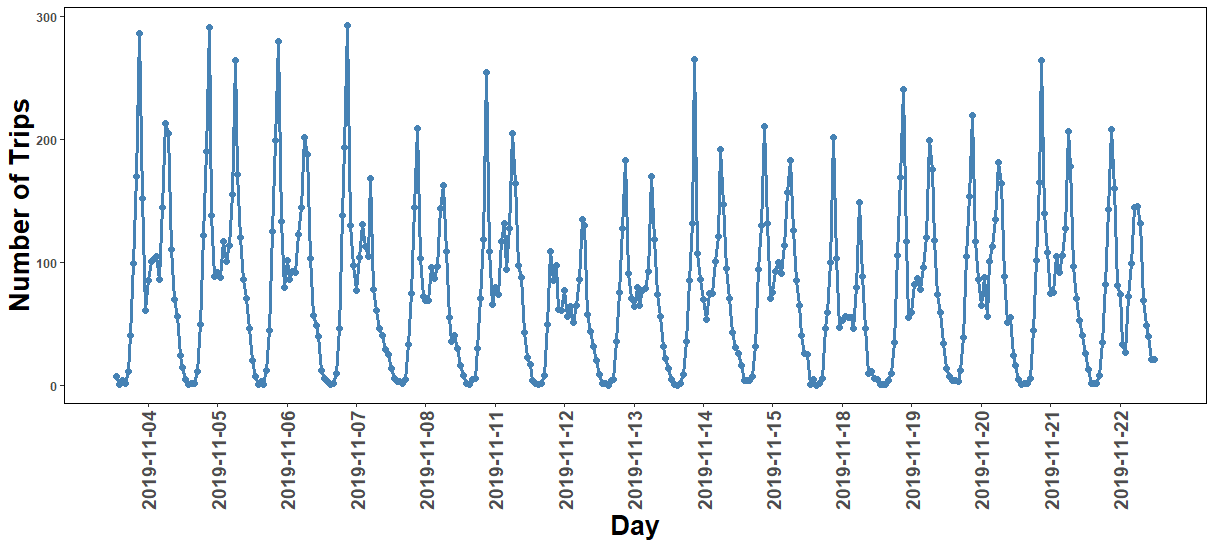}
\caption{The hourly number of trips between bike stations in the New York City Citi Bike sharing system for 15 weekdays in November 2019.   \label{fig::network_display}}
\end{figure}

To tackle the issue of non-Euclidean structure, numerous change point detection techniques have been developed for random objects. Some methods focus on detecting mean changes at a specific instant, such as those proposed in \cite{chen2015graph}, \cite{chen2017new}, \cite{dubey2020frechet}, and \cite{jiang2024two}, while others target distributional changes; see, for example, \cite{dubey2023change}, \cite{kanrar2024model}, and \cite{zhang2025change}. However, most existing approaches assume that, under the null hypothesis of no change point, the observations are identically distributed. This assumption often does not hold in real-world scenarios, e.g., when periodic effects are present. For example, dynamic transportation networks observed on an hourly basis are influenced by rush hour patterns. Figure \ref{fig::network_display} shows the hourly number of trips between bike stations in the New York City Citi Bike sharing system. Figure \ref{fig::network_display} indicates that these networks exhibit periodic behaviour, with peaks of network sizes at 8 am and 6 pm corresponding to morning and evening commuting surges. Suppose we aim to see whether the transportation networks change between days. Changes in the transportation network size due to rush hours would blur the changes one aims to detect. Analysing such data without accounting for periodic behaviour could obscure the detection of change points across days, as changes between hours might dominate.

Additional examples of periodic random objects appear in contexts such as monthly electricity generation and social communication networks. For electricity generation data, each monthly observation represents the percentage contribution of a source to net generation, introducing a natural 12-month period for the observations. For social communication networks, they always have time-varying graph structures where connections and message frequencies fluctuate periodically, such as daily usage cycles of internet traffic.

Motivated by these challenges, this paper proposes a novel procedure for detecting and localizing change points in sequences of random objects, with a dual focus on capturing the empirical distribution of the random objects and addressing periodic behaviour. The periodically distributed random objects are first grouped into blocks and a suitable nonparametric detector is then applied to identify the change block—that is, the block containing the change point. Given the detected block, a classification-based method is employed to localize the change point within this block. The proposed procedure is flexible and broadly applicable, accommodating a wide class of suitable nonparametric detectors for random objects, such as the ones using the metric distribution function, proposed by \cite{wang2023nonparametric}, and distance profiles, proposed by \cite{dubey2024metric}. To further study the theoretical properties of the proposed procedure, we consider a specific detector based on the metric distribution function mainly due to its natural extension to a product measure on the product metric space, which is essential for modelling periodic structures in random objects. Our method restricted to single change point detection offers rigorous type I error control, and guarantees change point estimation consistency under fixed and local alternatives. Notably, we relax the common assumption that the true change point must coincide with the boundary of periodic blocks—an unrealistic assumption frequently imposed in periodic change point detection literature for Euclidean data (see, e.g., \cite{guo2020nonparametric} and \cite{wang2022robust}). We further extend the proposed approach to multiple change point detection using the seeded binary segmentation algorithm (\citealt{kovacs2023seeded}), ensuring consistency in scenarios with multiple change points under fixed and local alternatives. From a theoretical perspective, the higher-order $U$-statistics in our detector lead to additional theoretical obstacles, marking a significant departure from the existing proof techniques employed in the area of $M$-estimation and change point estimation.

Our proposed approach represents a significant advance over existing methods, particularly in addressing periodic behaviour in random objects - a relatively unexplored area due to the lack of appropriate tools for capturing the relevant empirical distributions. Even for non-Euclidean data without periodic behaviour, our method remains competitive with leading nonparametric techniques, thanks to its comprehensive capture of random object distributions and minimal tuning parameter requirements.

Extensive simulations on periodic network data demonstrate the excellent finite-sample performance of our method. For random objects exhibiting periodic behaviour, our method demonstrates far superior power and accuracy in both detecting change points and pinpointing their locations. This is further validated through application to hourly transportation networks from the New York City Citi Bike sharing system, where the detected change points yield meaningful insights.

\section{Preliminaries and Methodologies}\label{sect::methodology single}

Let $Y_1,\ldots,Y_n$ denote an independent sequence of periodic random objects residing in a separable metric space $(\Omega,d)$. Recall that a subset of $\Omega$ is dense if its closure is $\Omega$; and $(\Omega,d)$ is separable if it has a countable dense subset; see \cite{wellner2013weak} for more details. For generic probability measures $P_l^{(i)}$ on $(\Omega,d)$ with $i\in\{1,\ldots,M\}$, consider the product measures $P_l^M=P_l^{(1)}\times\cdots\times P_l^{(M)}$, with $l=1,2$. Then $\{Y_i\}_{i=1}^n$ is defined to follow a $M$-periodic distribution $P_l^M$ if there exists a positive integer $M$ such that, for any integer $j\in\{1,\ldots,M\}$, (i) $Y_j\sim P_l^{(j)}$ and (ii) $Y_j,Y_{j+M},Y_{j+2M},\ldots,Y_{j+kM}$ have the same distribution for integer $k\geq 0$ satisfying $j+kM\leq n$.  We can see that $M=1$ corresponds to the case of independent and identically distributed observations. Additionally, define a function $r(j)$ by $r(j)=j+M-M\lfloor (j+M-1)/M \rfloor$, where $\lfloor\cdot\rfloor$ is the floor function. For any subsequence  $Y_{n_1},\ldots,Y_{n_2}$ with $1\leq n_1 < n_2\leq n$, $Y_{n_1},\ldots,Y_{n_2}$ is said to follow the periodic distribution $P_l^M$, denoted by $Y_{n_1},\ldots,Y_{n_2} \sim P_l^M$, if for any integer $j\in\{n_1,n_1+1,\ldots,n_1+M-1\}$, (i) $Y_{j}\sim P_l^{(r(j))}$, and (ii) $Y_j,Y_{j+M},Y_{j+2M},\ldots,Y_{j+kM}$ have the same distribution for integer $k\geq 0$ satisfying $j+kM\leq n_2$. The value of the period $M$ is often known and intuitive for the periodic sequences under consideration (\citealt{guo2020nonparametric,wang2022robust}). For example, the period of hourly transportation networks is 24, which can be observed in Fig. \ref{fig::network_display}. In the remainder of this paper, the period $M$ is assumed to be known.

Motivated by the methodology used in the Euclidean space from \cite{guo2020nonparametric}, the main strategy is to first arrange periodically distributed random objects into non-overlapping blocks of random objects, where the number of random objects in each block is equal to the period $M$. We then apply a scan statistic through almost the whole sequence of periodic blocks, and at each stage, partition the sequence of blocks into two contiguous segments: one preceding and one succeeding a candidate change block. Let $\bm{Z}_j=(Y_{(j-1)M+1},\ldots,Y_{jM})$ be a periodic block, for $j=1,\cdots,K_n$, where $K_n=\lfloor n/M \rfloor$ is the number of periodic blocks. We assume for simplicity that $n$ is a multiple of $M$. Otherwise, we can sample the missing random objects from the corresponding random objects in the previous periodic blocks or just ignore the remaining random objects. After the rearrangement, define $\bm{Z}_j\sim P_l^M$, meaning $(Y_{(j-1)M+1},\ldots,Y_{jM})\sim P_l^M$ for $j\in\{1,\ldots,K_n\}$ and $l\in\{1,2\}$.

Given different product probability measures $P_1^M$ and $P_2^M$, we are interested in testing the null hypothesis,
\begin{align}\label{formula::null hypothesis}
{\rm H}_0: &\bm{Z}_1,\bm{Z}_2,\ldots,\bm{Z}_{K_n}\sim P_1^M
\end{align}
against the single change point alternative,
\begin{align}\label{formula::alternative hypothesis}
{\rm H}_1: {\rm there~exists}~0< \tau <1~{\rm s.t.}
\begin{cases}
&\bm{Z}_1,\bm{Z}_2,\ldots,\bm{Z}_{\lfloor K_n\tau\rfloor}\sim P_1^M\\
&\bm{Z}_{\lfloor K_n\tau\rfloor + 1} \sim P_{1\oplus 2}^M\\
&\bm{Z}_{\lfloor K_n\tau\rfloor+2},\bm{Z}_{\lfloor K_n\tau\rfloor+3},\ldots,\bm{Z}_{K_n}\sim P_2^M,
\end{cases}
\end{align}
where $\bm{Z}_{\lfloor K_n \tau \rfloor + 1}$ is assumed to be the change block containing the true change point $Y_{\lfloor n \tau \rfloor + 1}$ and $P_{1\oplus 2}^M$ may differ from both $P_1^M$ and $P_2^M$. To see this, observe that $\bm{Z}_{\lfloor K_n \tau \rfloor + 1}=(Y_{\lfloor K_n \tau \rfloor M + 1},\cdots,Y_{(\lfloor K_n \tau \rfloor+1) M})$, when the change point is assumed to be located on the left boundary of the periodic block, that is, $Y_{\lfloor n \tau \rfloor + 1}$ locates at the left boundary of the change block $\bm{Z}_{\lfloor K_n \tau \rfloor + 1}$ with $\lfloor n \tau \rfloor=\lfloor K_n \tau \rfloor M$, we always have $\bm{Z}_{\lfloor K_n\tau\rfloor + 1}\sim P_2^M$ under ${\rm H}_1$  and thus $P_{1\oplus 2}^M=P_2^M$. In this case, the change point detection problem is analogous to the one with the independent and identically distributed assumptions. Although this assumption that the change point locates at the left boundary of the periodic block is commonly assumed in the nonparametric change point detection literature, see \cite{guo2020nonparametric} and \cite{wang2022robust}, it is too restricted for the periodic data.  When the true change point does not align with the left boundary of the changed periodic block, $Y_{\lfloor K_n \tau \rfloor M + 1},\cdots, Y_{\lfloor n\tau\rfloor} \sim P_1^M$ while  $Y_{\lfloor n\tau\rfloor+1},\cdots, Y_{(\lfloor K_n \tau \rfloor+1) M} \sim P_2^M$, and thus $P_{1\oplus 2}^M$ is neither $P_1^M$ nor $P_2^M$. The irregular situation of the change block $\bm{Z}_{\lfloor K_n\tau\rfloor + 1} \sim P_{1\oplus 2}^M$ poses significant challenges. To deal with these issues, a novel change point detection procedure is developed in the next section.

\subsection{General Detection Procedure for Periodic Random Objects}\label{subsect::two step change point}

After arranging periodically distributed random objects into periodic blocks, we start with a general change point detector, see, e.g.,  \cite{kojadinovic2021nonparametric}, given by
\begin{align}\label{eq::test statistic}
\hat{B}_n=\max_{u\in \mathcal{C}_b} K_n \hat{B}_n(u).
\end{align}
based on a general two-sample test statistic $\hat{B}_n(u)$ and $\mathcal{C}_b=[b,1-b]\subset [0,1]$.  As the scan statistic $\hat{B}_n$ relies on a two-sample test statistic $\hat{B}_n(u)$, an assumption is needed to ensure the asymptotic theory hold such as $\tau$ belongs to a compact interval $\mathcal{C}_b=[b,1-b]\subset [0,1]$, for some positive constant $b<1$. This is a common assumption as the detection of a change block is quite challenging when the change block is close to the first or the last observation. Here, the statistic $\hat{B}_n(u)$ is a general two-sample test statistic and is required to measure the distributional discrepancy between $\bm{Z}_1,\ldots,\bm{Z}_{\lfloor K_nu\rfloor}$ and $\bm{Z}_{\lfloor K_nu\rfloor+1},\ldots,\bm{Z}_{K_n}$. Suppose we have two functionals, $\hat{f}_1(\bm{Z}_1,\ldots,\bm{Z}_{\lfloor K_nu\rfloor})$ and $\hat{f}_2(\bm{Z}_{\lfloor K_nu\rfloor+1},\ldots,\bm{Z}_{K_n})$, which represent of the distribution of the first and second data segments, respectively. Then, a general Cramér–von Mises–type two-sample statistic can be expressed as
\[
\hat{B}_n(u) = \frac{\lfloor K_n u\rfloor(K_n-\lfloor K_n u\rfloor)}{K_n^2} \left\{ \hat{f}_1(\bm{Z}_1,\ldots,\bm{Z}_{\lfloor K_nu\rfloor}) - \hat{f}_2(\bm{Z}_{\lfloor K_nu\rfloor+1},\ldots,\bm{Z}_{K_n}) \right\}^2
\]
For instance, $\hat{B}_n(u)$ can be chosen as the statistic in (\ref{formula::MDF two sample test}) later.

Assume we have the limiting distribution $\mathcal{L}$ of the general detector $\hat{B}_n$ under the null ${\rm H}_0$ in (\ref{formula::null hypothesis}), one should propose to reject ${\rm H}_0$ if $\rho\leq \alpha$ for a level $\alpha$ test, where $\rho=pr_{{\rm H}_0}(\mathcal{L}\geq \hat{B}_n)$ is the asymptotic $p$-value of the test. However, the distribution of $\mathcal{L}$ may depend on the unknown periodic data distribution, making the exact rejection threshold intractable. We can consider a random permutation procedure to approximate the $p$-value which is described below, see, e.g., \cite{chung2016asymptotically}, \cite{dubey2023change} and \cite{dubey2024metric}.


Define $J_{K_n}$ to be the set of $K_n!$ permutations of $\{1,\ldots,K_n\}$. We first randomly sample permutations $\pi_1,\ldots,\pi_{L_{\rm perm}}$ from $J_{K_n}$ with replacement, where $\pi_l=\{\pi_l(1),\ldots,\pi_l(K_n)\}$ for $l=1,\ldots,L_{\rm perm}$ and $L_{\rm perm}$ is the number of randomly selected permutations. Then let $\pi_0=(\pi_0(1),\ldots,\pi_0(K_n))$ be the ordered set such that $\pi_0(j)=j$ for $j=1,\ldots,K_n$. For each $l=1,\ldots,{L_{\rm perm}}$, let $\hat{B}_n^{\pi_l}$ be the single change point detector evaluated on the reordering data given by $\bm{Z}_{\pi_l(1)},\ldots,\bm{Z}_{\pi_l(K_n)}$. The $p$-value of the permutation test is finally calculated as
$\hat{\rho}_{L_{\rm perm}}^s=\sum_{l=0}^{L_{\rm perm}}\mathbb{I}\{ \hat{B}_n^{\pi_l} \geq \hat{B}_n \}/(L_{\rm perm}+1)$ where $\mathbb{I}(\cdot)$ is the indicator function, and the single change point hypothesis test is rejected when $\hat{\rho}_{L_{\rm perm}}^s\leq \alpha$ for a level $\alpha$ test. By \cite{chung2016asymptotically}, one can see that for a sufficiently large $L_{\rm perm}$, $\hat{\rho}_{L_{\rm perm}}^s$ controls the type I error of the test reasonably well under ${\rm H}_0$. Additionally, in practice, we consider a certain premature termination proposed in \cite{hapfelmeier2023efficient} to improve the efficiency of the permutation test. At stage $l\leq L_{\rm perm}$, we accept ${\rm H}_0$ when
$\hat{\rho}_l^s \geq \{\log (0.1) + l\log (47/48)\}/ \log ( 47/72)$ and reject ${\rm H}_1$ when
$\hat{\rho}_l^s \leq \{\log (10) + l\log (47/48) \}/ \log ( 47/72 )$.
The choice of this early stopping criterion has been shown to maintain a nominal significance level in \cite{hapfelmeier2023efficient} and in our simulation study in \S \ref{sect::simulation}.

If the null hypothesis ${\rm H}_0$ in (\ref{formula::null hypothesis}) is rejected based on the block-based detector $\hat{B}_n$ using the random permutation procedure, the estimated change block is $\bm{Z}_{\lfloor K_n\hat{\tau}_b \rfloor+1}$ where 
\begin{align}\label{formula:: def of change point estimation}
    \hat{\tau}_b = \argmax_{u\in\mathcal{C}_b} \hat{B}_n(u).
\end{align}
Given a detected change block $\bm{Z}_{\lfloor K_n\hat{\tau}_b \rfloor+1}$, though optional, it is sometimes of interest to pinpoint the location of the change point within that block. To this end, for $m\in\{1,2,\ldots,M\}$, suppose we have a general marginal two-sample test statistic $\hat{S}_n^{(m)}$ 
to assess whether the distributions of two samples $\{Z_{1,m},\ldots,Z_{\lfloor K_n\hat{\tau}_b\rfloor,m}\}$ and $\{Z_{\lfloor K_n\hat{\tau}_b\rfloor+2,m},\ldots,Z_{K_n,m}\}$ are statistically identical for each $m\in\{1,2,\ldots,M\}$, where $Z_{i,m}$ denotes the $m$-th element of the block $\bm{Z}_i$ that is, $Z_{i,m}=Y_{(i-1)M+m}$. Let $P_{1,s}^{(m)}$ and $P_{2,s}^{(m)}$ be the Borel probability measures corresponding to $\{Z_{1,m},\ldots,Z_{\lfloor K_n\hat{\tau}_b\rfloor,m}\}$ and $\{Z_{\lfloor K_n\hat{\tau}_b\rfloor+2,m},\ldots,Z_{K_n,m}\}$, respectively. The corresponding null and alternative hypotheses for the marginal test are
\begin{align}
    {\rm H}_{0,m}^s: P_{1,a}^{(m)} = P_{2,a}^{(m)}\label{marginal test null}\\
    {\rm H}_{1,m}^s: P_{1,a}^{(m)} \neq P_{2,a}^{(m)} \label{marginal test alter}.
\end{align}

We then construct the set $\mathcal{V}_n$ consisting of indices $m$ for which the test statistic $\hat{S}_n^{(m)}$ leads to the rejection of the null hypothesis ${\rm H}_{0,m}^s$ given in (\ref{marginal test null}). For each $\nu\in \mathcal{V}_n$, we apply a nearest neighbour classification method to determine whether the observation $Z_{\lfloor K_n\hat{\tau}_b\rfloor+1,\nu}$ is more likely to belong to distribution $P_{1,a}^{(\nu)}$ or $P_{2,a}^{(\nu)}$. Specifically, for a tuning parameter $e_{nn}$ which represents the number of the nearest neighbours, let $\hat{d}_k=d(Z_{k,\nu},Z_{\lfloor K_n\hat{\tau}_b\rfloor+1,\nu})$ for $k=1,\ldots,\lfloor K_n\hat{\tau}_b\rfloor,\lfloor K_n\hat{\tau}_b\rfloor+2,\ldots,K_n$. Write $\hat{d}_{(1)}\leq\hat{d}_{(2)}\leq \cdots \leq \hat{d}_{(K_n-1)}$ for the ordered values of $\{\hat{d}_k\}_{k=1}^{K_n-1}$. Let $r=\hat{d}_{(e_{nn})}$, $C_{1,\nu}$ and $C_{2,\nu}$ be the cardinalities of the sets $\{k: d(Z_{k,\nu},Z_{\lfloor K_n\hat{\tau}_b\rfloor+1,\nu})\leq r, k=1,\ldots,\lfloor K_n\hat{\tau}_b\rfloor\}$ and $\{k: d(Z_{k,\nu},Z_{\lfloor K_n\hat{\tau}_b\rfloor+1,\nu})\leq r, k=\lfloor K_n\hat{\tau}_b\rfloor+2,\ldots,K_n\}$, respectively. If $C_{1,\nu}>C_{2,\nu}$ we classify $Z_{\lfloor K_n\hat{\tau}_b\rfloor+1,\nu}$ as belonging to $P_{1,a}^{(\nu)}$ and denote the classification result as $\hat{h}^{(\nu)}=1$. If $C_{1,\nu}<C_{2,\nu}$, $Z_{\lfloor K_n\hat{\tau}_b\rfloor+1,\nu}$ is assigned to $P_{2,a}^{(\nu)}$ and let $\hat{h}^{(\nu)}=2$. For $\nu\in \mathcal{V}_n$, if there exists some $\nu$ such that $Z_{\lfloor K_n\hat{\tau}_b\rfloor+1,\nu}$ is classified as belonging to $P_{2,a}^{(\nu)}$, let $\hat{\nu}$ be the smallest index; otherwise, let $\hat{\nu}=M+1$. Here, $\hat{\nu}$ can be viewed as the within-block index and 
the performance of the classification approach considered above is supported by the universal consistency of the nearest neighbour classification in metric spaces, see e.g., Theorem 1 in \cite{chaudhuri2014rates}, given the tuning parameter $e_{nn}$ satisfying $e_{nn}/n\to0$ and $e_{nn}\to\infty$ as $n\to\infty$.

We now explain the necessity of testing ${\rm H}_{0,m}^s$ given by (\ref{marginal test null}) in the proposed procedure. Observe that under ${\rm H}_1$ in (\ref{formula::alternative hypothesis}), $P_1^M\neq P_2^M$, which implies that there exists some $m\in\{1,\ldots,M\}$ such that $P_1^{(m)}\neq P_2^{(m)}$. However, it is also possible that for some values of $m$, the marginal distributions satisfy $P_1^{(m)}=P_2^{(m)}$. This presents a key challenge in change point detection for periodic data as we do not know in advance which indices $m$ such that $P_1^{(m)}\neq P_2^{(m)}$. Therefore, it is essential to test ${\rm H}_{0,m}^s$ for each $m$ within the detected change block. Once the change block is consistently identified, the marginal tests are effective in picking the indices $m$ which exhibit distributional changes, enhancing the precision of change point localization within the periodic block.

The final detected change point location, according to the proposed procedure is 
\begin{align}\label{formula::two step change point location output}
    \hat{l}_F = \lfloor K_n\hat{\tau}_b \rfloor M + \hat{\nu}.
\end{align} 
And the estimation of $\tau$ in this general procedure is calculated as 
\begin{align}\label{formula::final change point est}
    \hat{\tau}_F =  \hat{l}_F/n.
\end{align}

The above procedure for a single change point can be extended to multiple change point detection for periodic random objects, which is developed in \S S1 of the supplementary material.

\subsection{Metric-Distribution-Function-Based Detector}
As discussed above, the change block $\bm{Z}_{\lfloor K_n\tau\rfloor + 1}$ follows neither $P_1^M$ nor $P_2^M$. The irregular situation of the change block poses additional theoretical challenges as the commonly used assumption in the literature that $\bm{Z}_{\lfloor K_n\tau\rfloor + 1},$  $\ldots,\bm{Z}_{K_n}$ are independent and identically distributed does not hold. Therefore, to study the theoretical properties of the general procedure developed in \S \ref{subsect::two step change point}, we need to consider a specific detector which is constructed based on the metric distribution function.

The metric distribution function, introduced in \cite{wang2023nonparametric}, is a powerful tool for quantifying the variability of random objects, while an analogous approach, using distance profiles, is developed in \cite{dubey2024metric}. Let $(G, \mathcal{G}, \mathbb{P})$ be a probability space where $\mathcal{G}$ denotes the Borel $\sigma$-algebra on the set $G$ and $\mathbb{P}$ is a probability measure. For a random object $X$ located in a separable metric space $\Omega$ with a metric $d$, it is defined as a measurable map $X: G \to \Omega$. This mapping induces a probability measure $P$ on $\Omega$, given by $P(S) = \mathbb{P}(X^{-1}(S))$ for every Borel measurable set $A \subseteq \Omega$. Let $\bar{B}(w,r)=\{v:d(w,v)\leq r\}$ be a closed ball with centre $w$ and radius $r\geq 0$, and write $\delta(w,v,x)=\mathbb{I}\{x\in\bar{B}(w,d(w,v))\}$. For any points $w,v\in\Omega$, the metric distribution function is given by
\[
F_P(w,v)=pr(X\in\bar{B}(w,d(w,v)))=E_{X\sim P}\left[ \delta(w,v,X) \right].
\]
Given a sequence of independent and identically distributed observations $X_1,\ldots,X_n$ from $P$, the metric distribution function is estimated as
$\hat{F}(w,v)=n^{-1}\sum_{i=1}^n \delta(w,v,X_i)$.
\cite{wang2023nonparametric} show that the metric distribution function has one-to-one correspondence with a probability measure under a specific condition on $(\Omega,d)$, that is, when the metric $d$ is directionally $(\epsilon,\varsigma,L)$-limited at the support of the probability measure $P$ (\citealt{federer2014geometric,wang2023nonparametric}). See the Appendix for further discussion of the concept of a directionally $(\epsilon,\varsigma,L)$-limited metric.

Suppose we have data objects from two unknown Borel probability measures $P_1$ and $P_2$ on a metric space $(\Omega,d)$ with equal sample sizes. Let $F_{P_k}(w,v)$ be the metric distribution functions for $P_k$,    $k=1,2$. The Cram\'{e}r-von Mises-like statistic, see \cite{wang2023nonparametric}, to test ${\rm H}_0:P_1=P_2$ is given by 
\[
{\rm MCVM}(P_1,P_2) = \sum_{k=1}^2\int_{(w,v)\in\Omega\times \Omega} \left( F_{P_1}(w,v) - F_{P_2}(w,v) \right)^2 d P_k(w)d P_k(v).
\]

A consistent estimator of this population test statistic ${\rm MCVM}(P_1,P_2)$ is obtained via the empirical metric distribution functions which are the consistent estimators of the corresponding metric distribution functions $F_{P_1}(w,v)$ and $F_{P_2}(w,v)$. 


Given the natural extension of the metric distribution function to a product measure on the product metric space, let $\hat{F}_{j:k}$ be the empirical joint metric distribution function computed from the sequence $\bm{Z}_j,\ldots,\bm{Z}_k$ of available observations. More formally, let $\Omega^M$ be the Cartesian product space of $M$ copies of $\Omega$. F+or any integers $j,k\geq 1$, $\bm{w}=(w_1,\ldots,w_M)$, and $\bm{v}=(v_1,\ldots,v_M)$ with $\bm{w},\bm{v}\in\Omega^M$, we have
\begin{align*}
\hat{F}_{j:k}(\bm{w},\bm{v})=\begin{cases}
\frac{1}{k-j+1}\sum_{i=j}^k \delta(\bm{w},\bm{v},\bm{Z}_i) ,~~&{\rm if}~j\leq k\\
0,&{\rm otherwise},
\end{cases}
\end{align*}
where
\[
\delta(\bm{w},\bm{v},\bm{Z}_i)=\prod_{m=1}^M \mathbb{I}\left\{ Y_{(i-1)M+m} \in\bar{B}(w_m,d(w_m,v_m)) \right\}.
\]
Define $\hat{\mathcal{B}}^{(1)}_n(u)$ and $\hat{\mathcal{B}}^{(2)}_n(u)$ as
\begin{align}\label{eq::D1}
\hat{\mathcal{B}}^{(1)}_n(u)=\frac{1}{\left(\lfloor K_n u\rfloor\right)^2} \sum_{i,j\in \mathcal{K}_u}  W(\bm{Z}_i,\bm{Z}_j) \left\{ \hat{F}_{1:\lfloor K_n u\rfloor}(\bm{Z}_i,\bm{Z}_j) - \hat{F}_{(\lfloor K_n u\rfloor+1):K_n}(\bm{Z}_i,\bm{Z}_j) \right\}^2 , 
\end{align}
and 
\begin{align}\label{eq::D2}
\hat{\mathcal{B}}^{(2)}_n(u)=\frac{1}{\left(K_n-\lfloor K_n u\rfloor\right)^2} \sum_{i,j\in \mathcal{K}_{-u}}  W(\bm{Z}_i,\bm{Z}_j) \left\{ \hat{F}_{1:\lfloor K_n u\rfloor}(\bm{Z}_i,\bm{Z}_j) - \hat{F}_{(\lfloor K_n u\rfloor+1):K_n}(\bm{Z}_i,\bm{Z}_j) \right\}^2 , 
\end{align}
where $W(\cdot , \cdot)$ is a non-negative and bounded weight function, $\mathcal{K}_u=\{1,2,\ldots,\lfloor K_n u\rfloor\}$ and $\mathcal{K}_{-u}=\{\lfloor K_n u\rfloor+1,\lfloor K_n u\rfloor+2,\ldots,K_n\}$. The statistic $\hat{B}_n(u)$ is then
\begin{align}\label{formula::MDF two sample test}
\hat{B}_n(u)=\frac{\lfloor K_n u\rfloor(K_n-\lfloor K_n u\rfloor)}{K_n^2} \left( \hat{\mathcal{B}}^{(1)}_n(u) + \hat{\mathcal{B}}^{(2)}_n(u)  \right).
\end{align}
Notably, to enhance flexibility, we introduce the weights $W(\bm{w},\bm{v})$ shown in (\ref{eq::D1}) and (\ref{eq::D2}) into test statistic $\hat{B}_n(u)$. We allow for these data adaptive weights can be tuned appropriately to enhance the detection capacity of the test statistic in theory. In our numerical studies, we only consider the special case when all the weights are set to be 1. The specific change point detector is constructed via substituting (\ref{formula::MDF two sample test}) into (\ref{eq::test statistic}), that is, $\hat{B}_n=\max_{u\in \mathcal{C}_b} K_n \hat{B}_n(u).$

Similarly, we can borrow the marginal two-sample test in \cite{wang2023nonparametric} to construct the test statistic $\hat{S}_n^{(m)}$, $m\in\{1,\ldots,M\}$ for the null hypothesis ${\rm H}_{0,m}^s$ in (\ref{marginal test null}) and the alternative ${\rm H}_{1,m}^s$ in (\ref{marginal test alter}). The corresponding test statistic is given by
\begin{align*}
    \hat{S}_n^{(m)} = \frac{\lfloor K_n\hat{\tau}_b\rfloor(K_n-\lfloor K_n\hat{\tau}_b\rfloor-1)}{K_n^2} \left( \hat{\mathcal{S}}^{(1,m)}_n + \hat{\mathcal{S}}^{(2,m)}_n  \right),
\end{align*}
where
\begin{align*}
\hat{\mathcal{S}}^{(1,m)}_n=\frac{1}{\left(\lfloor K_n\hat{\tau}_b\rfloor\right)^2} \sum_{i,j\in\mathcal{K}_{\hat{\tau}_b}}  W(Z_{i,m},Z_{j,m}) \left\{ \hat{F}_{1:\lfloor K_n\hat{\tau}_b\rfloor}(Z_{i,m},Z_{j,m}) - \hat{F}_{(\lfloor K_n\hat{\tau}_b\rfloor+2):K_n}(Z_{i,m},Z_{j,m}) \right\}^2 , 
\end{align*}
and 
\begin{align*}
\hat{\mathcal{S}}^{(2,m)}_n=&\frac{1}{\left(K_n-\lfloor K_n\hat{\tau}_b\rfloor-1\right)^2} \sum_{i,j\in\widetilde{\mathcal{K}}_{-\hat{\tau}_b}} W(Z_{i,m},Z_{j,m}) \\
&\times\left\{ \hat{F}_{1:\lfloor K_n\hat{\tau}_b\rfloor}(Z_{i,m},Z_{j,m}) - \hat{F}_{(\lfloor K_n\hat{\tau}_b\rfloor+2):K_n}(Z_{i,m},Z_{j,m}) \right\}^2 , 
\end{align*}
with $\widetilde{\mathcal{K}}_{-\hat{\tau}_b}=\{\lfloor K_n\hat{\tau}_b\rfloor +2,\lfloor K_n\hat{\tau}_b\rfloor +3,\ldots,K_n\}$.

\section{Theoretical Properties}\label{sect::theory}


Given the specific detector $\hat{B}_n$ constructed by (\ref{formula::MDF two sample test}) and (\ref{eq::test statistic}), we are now able to derive the theoretical results for the single change point detection. The theoretical results cover the limiting distribution of the change point detector under the null, the power analysis of the detector under the local alternatives and the convergence rate of the detected change point location under the alternative. In Theorem \ref{thm::asymptotics under null} below we give the asymptotic distribution of $\hat{B}_n$ under the null hypothesis.

\begin{theorem}\label{thm::asymptotics under null}
For a specific detector $\hat{B}_n$  constructed by (\ref{formula::MDF two sample test}) and (\ref{eq::test statistic}), under ${\rm H}_0$ in (\ref{formula::null hypothesis}) and Assumptions \ref{assump 1}-\ref{assump 4} in the Appendix, as $n\to\infty$, $\hat{B}_n$ converges in distribution to  $\mathcal{L}=\sup_{u\in\mathcal{C}_b}\sum_{j=1}^\infty \lambda_j \mathcal{W}_j^2(u) $, where $\lambda_j$ are constants depending on $P_1^M$ in (\ref{formula::null hypothesis}) and $\mathcal{W}_1,\mathcal{W}_2,\ldots$ are independent and identically distributed Gaussian processes with zero mean function and autocovariance function $\Sigma(u_1,u_2) = f(u_1,u_2)\min(u_1,u_2) + \min(1-u_1,1-u_2)/f(u_1,u_2)$, and $f(u_1,u_2)=\sqrt{(1-u_1)(1-u_2)/(u_1u_2)}$.
\end{theorem}

\begin{remark}
    The specific detector $\hat{B}_n$ is constructed by combining the scan technique with the two-sample test statistic proposed in \cite{wang2023nonparametric}. The limiting distribution of the two-sample test statistic in \cite{wang2023nonparametric} takes the form $\sum_{j=1}^\infty \lambda_j \mathcal{W}_j^2(u_f)$, where $u_f$ is fixed and depends on the sample sizes of the two samples. The metric-distribution-function-based test statistic is a $U$-statistic, and the main theoretical tool used in the proof is a projection technique for $U$-statistics. A similar result and analogous projection techniques for $U$-processes, albeit in a different context for non-perodic data, can be found in the two-sample test and change point detection literature such as \cite{dubey2023change} and \cite{dubey2024metric}.
\end{remark}

Theorem \ref{thm::asymptotics under null} shows that the detector $\hat{B}_n$ in (\ref{eq::test statistic}) converges to the supremum of a weighted infinite sum of independent and identically distributed squared Gaussian processes under ${\rm H}_0$. To derive the asymptotic distribution under the null in Theorem \ref{thm::asymptotics under null}, the detector $\hat{B}_n$ is split into several parts following the projection technique for $U$-statistics. By utilizing L\'{e}vy-type maximal inequalities and decoupling results for $U$-statistics (\citealt{de1995decoupling,eichelsbacher2001moderate}), some of parts are asymptotically negligible while some of parts determine the asymptotic null distribution.


We now turn to the analysis of the test power against local alternatives. 
To define a sequence of local alternatives, a distance measure between the distributions $P_1^M$ and $P_2^M$, denoted as $\Delta=\Delta(P_1^M,P_2^M)$, is essential, which has been established in \cite{pan2020ball,wang2023nonparametric} as
\begin{align*}
\Delta=&E_{\bm{Z},\widetilde{\bm{Z}}\sim P_1^M} \left\{ W(\bm{Z},\widetilde{\bm{Z}}) \left( F^{(1)}(\bm{Z},\widetilde{\bm{Z}})-F^{(2)}(\bm{Z},\widetilde{\bm{Z}})  \right)^2 \right\} \\
&+ E_{\bm{Z},\widetilde{\bm{Z}}\sim P_2^M} \left\{ W(\bm{Z},\widetilde{\bm{Z}})  \left( F^{(1)}(\bm{Z},\widetilde{\bm{Z}})-F^{(2)}(\bm{Z},\widetilde{\bm{Z}})  \right)^2 \right\},
\end{align*}
where $F^{(1)}(\bm{Z},\widetilde{\bm{Z}})=E_{\bm{Z}^\prime}(\delta(\bm{Z},\widetilde{\bm{Z}},\bm{Z}^\prime))$ with $\bm{Z}^\prime\sim P_1^M$ and $F^{(2)}(\bm{Z},\widetilde{\bm{Z}})=E_{\bm{Z}^{\prime\prime}}(\delta(\bm{Z},\widetilde{\bm{Z}},\bm{Z}^{\prime\prime}))$ with $\bm{Z}^{\prime\prime}\sim P_2^M$ and both $\bm{Z}^\prime$ and $\bm{Z}^{\prime\prime}$ are independent of $\bm{Z},\widetilde{\bm{Z}}$. Note that $\Delta$ corresponds to a special case of the homogeneity test statistic introduced in \cite{wang2023nonparametric}, whose results imply that, under mild conditions, $\Delta=0$ if and only if $P_1^M=P_2^M$. Therefore, $\Delta$ can be used to measure the distance between ${\rm H}_0$ in (\ref{formula::null hypothesis}) and ${\rm H}_1$ in (\ref{formula::alternative hypothesis}).

Given the divergence measure $\Delta$, the sequence of local alternatives ${\rm H}_{1,n}$ that shrinks to ${\rm H}_0$ can be defined as
\begin{align}\label{formula::contiguous alternatives}
{\rm H}_{1,n}=\{ (P_1^M,P_2^M):\Delta=a_n \}
\end{align}
with $a_n\to 0$ and $\lfloor n\tau\rfloor(n-\lfloor n\tau\rfloor)a_n/n\to\infty$ as $n\to\infty$. For the first case where the limiting distribution $\mathcal{L}$ is assumed to be known, the power of the level $\alpha$ test is defined as $\beta_n^\alpha = pr_{{\rm H}_{1,n}}(\rho\leq \alpha)$ recalling that $\rho$ is the asymptotic $p$-value. Here, the asymptotic $p$-value depends on $\mathcal{L}$ in Theorem \ref{thm::asymptotics under null}. For the second case for the permutation test, we estimate the asymptotic $p$-value by $\hat{\rho}_{L_{\rm perm}}^s$ and thus the power is evaluated as $\widetilde{\beta}_n^\alpha = pr_{{\rm H}_{1,n}}(\hat{\rho}_{L_{\rm perm}}^s\leq \alpha)$. Theorem \ref{thm::asymptotics under alternative} shows 
that the power of the test converges to one as $n,L_{\rm perm} \to \infty$ under a sequence of local alternatives, provided that $n a_n \to \infty$. This condition ensures that the deviation from the null does not vanish too rapidly with increasing sample size. 

\begin{theorem}\label{thm::asymptotics under alternative}
For a specific detector $\hat{B}_n$  constructed by (\ref{formula::MDF two sample test}) and (\ref{eq::test statistic}), under Assumptions \ref{assump 1}-\ref{assump 4} in the Appendix, for any $\alpha\in(0,1)$ and for any sequence of alternatives given by ${\rm H}_{1,n}$  satisfying (\ref{formula::contiguous alternatives}), the power of the oracle test $\beta_n^\alpha\to 1$ as $n\to\infty$ while the power of the permutation-based test $\widetilde{\beta}_n^\alpha\to 1$ as $n,L_{\rm perm}\to\infty$.
\end{theorem}

\begin{remark}
    Theorem \ref{thm::asymptotics under alternative} is highly dependent on the power analysis of the metric-distribution-function-based two-sample test statistic, a point not discussed in  \cite{wang2023nonparametric}. When there is no periodic pattern, i.e., $M=1$, $\bm{Z}_{\lfloor K_n\tau\rfloor + 1},\ldots,\bm{Z}_{K_n}$ are independent and identically distributed, thus in this setting, an analogous result for the test power, though arising in a different context, can be found in the two-sample test and change point detection literature such as \cite{erlemann2022cramer,dubey2023change} and \cite{dubey2024metric}.
\end{remark}


Theorem \ref{thm::asymptotics under alternative} is established under the alternatives where the change block $\bm{Z}_{\lfloor K_n\tau\rfloor + 1}$ follows neither $P_1^M$ nor $P_2^M$. The irregular situation of the change block poses additional theoretical challenges as the commonly used assumption in the literature that $\bm{Z}_{\lfloor K_n\tau\rfloor + 1},$  $\ldots,\bm{Z}_{K_n}$ are independent and identically distributed does not hold. The same situation happens in the change point estimation in the following.




To study the consistency of the change point estimation $\hat{\tau}_F$, we consider two cases when (i) the alternative is fixed, and (ii) the alternative  $(P_1^M,P_2^M)\in{\rm H}_{1,n}$ with ${\rm H}_{1,n}$ defined in (\ref{formula::contiguous alternatives}). 
Borrowing ideas from projection techniques for $U$-statistics (\citealt{wang2023nonparametric}) by constructing $B_n(u)$ as
\begin{align*}
B_n(u)=\eta(u)\zeta(u,\tau)\left( \mathcal{B}^{(1)}_n(u)+ \mathcal{B}^{(2)}_n(u)   \right) 
\end{align*}
where $\eta(u)=u(1-u)$, $\zeta(u,\tau) = (1-\tau)^2\mathbb{I}(u\leq \tau)/(1-u)^2 +\tau^2\mathbb{I}(u>\tau)/u^2$,
\[
\mathcal{B}^{(1)}_n(u)=\frac{1}{(\lfloor K_n u\rfloor)^2}\sum_{i,j\in \mathcal{K}_u}W(\bm{Z}_i,\bm{Z}_j)\left( F^{(1)}(\bm{Z}_i,\bm{Z}_j) - F^{(2)}(\bm{Z}_i,\bm{Z}_j) \right)^2,
\]
and
\[
\mathcal{B}^{(2)}_n(u)  =\frac{1}{(K_n-\lfloor K_n u\rfloor)^2}\sum_{i,j\in \mathcal{K}_{-u}}W(\bm{Z}_i,\bm{Z}_j)\left( F^{(1)}(\bm{Z}_i,\bm{Z}_j) - F^{(2)}(\bm{Z}_i,\bm{Z}_j) \right)^2.
\]
The primary challenge arises from the fact that $\tau$ cannot be guaranteed to be the maximum of $B_n(u)$, meaning that the inequality $B_n(\tau) - B_n(\hat{\tau}_F) > 0$ does not always hold. This is an uncommon situation in the area of $M$-estimation and in change point detection,  marking another significant departure from the proof techniques employed in classical change point estimation for random objects, such as \cite{dubey2023change}. To address this critical challenge, we develop a new theorem below. To the best of our knowledge, Theorem \ref{lemma::bound of detector} has no counterpart in the change point literature.

\begin{theorem}\label{lemma::bound of detector}
For a specific detector $\hat{B}_n$  constructed by (\ref{formula::MDF two sample test}) and (\ref{eq::test statistic}), under Assumptions \ref{assump 1}-\ref{assump 4} in the Appendix, for some positive constants $C_1$ and $C_2$, there exist a sequence of events $\{\mathcal{A}_n\}_{n=1}^\infty$ with $pr(\mathcal{A}_n)\to 1$ as $n\to\infty$ and an $n_0$ such that, for each $n \geq n_0$ 
\[
C_1|\hat{\tau}_F-\tau|\Delta \leq\left| \hat{B}_n(\hat{\tau}_F)- \hat{B}_n(\tau) +  B_n(\tau)-B_n(\hat{\tau}_F)\right| \leq C_2|\hat{\tau}_F-\tau|\Delta
\]
whenever the event $\mathcal{A}_n$ occurs.
\end{theorem}

Theorem \ref{lemma::bound of detector} ensures that
\[
pr\left( |\hat{\tau}_F-\tau|\geq C_3 , \mathcal{A}_n \right)\leq pr\left( \left| \hat{B}_n(\hat{\tau}_F)- \hat{B}_n(\tau) +  B_n(\tau)-B_n(\hat{\tau}_F)\right|\geq C_1C_3\Delta , \mathcal{A}_n \right)+o(1)
\]
holds for any constant $C_3$ and some event $\mathcal{A}_n$ with $pr(\mathcal{A}_n)\to 1$ as $n\to\infty$. This, together with Markov's inequality and Lemma S4 in the supplementary material on the convergence rates for 
\[
E\left( \sup_{|u-\tau|\leq \iota_n} \left| \hat{B}_n(u)-\hat{B}_n(\tau) +  B_n(\tau)-B_n(u) \right|  \right)
\]
where $\iota_n\to 0$ and $n\iota_n\to \infty$ as $n\to\infty$, enables us to derive the consistency of $\hat{\tau}_F$ in Theorem \ref{thm::single cp estimation}.

\begin{theorem}\label{thm::single cp estimation}
For a specific detector $\hat{B}_n$ constructed by (\ref{formula::MDF two sample test}) and (\ref{eq::test statistic}), under Assumptions \ref{assump 1}-\ref{assump 4} in the Appendix, for any fixed alternative $P_1^M\neq P_2^M$, $|\hat{\tau}_F-\tau|=O_p( \gamma_n)$ as $n\to\infty$ for any $\gamma_n$ asymptotically larger than $n^{-1}$.
For local alternatives ${\rm H}_{1,n}$ given in (\ref{formula::contiguous alternatives}), there exists some constant $C>0$ such that 
$pr_{{\rm H}_{1,n}}(na_n |\hat{\tau}_F-\tau| \geq C) \to 0$,
as $n\to\infty$.
\end{theorem}

Theorem \ref{thm::single cp estimation} provides the asymptotic near-optimal convergence rate for change point estimation under the fixed alternative scenario while change point estimation is consistent in the local alternative scenario when $na_n\to\infty$ as $n\to \infty$. Theorem \ref{thm::single cp estimation} relates to classical  convergence rate results in the area of nonparametric change point detection, see, e.g., \cite{dumbgen1991asymptotic} and \cite{dubey2023change}. 

Theoretical results of the multiple change point detection for periodic random objects can be found in \S S1 of the supplementary material.

\section{Simulation}\label{sect::simulation}

We evaluate the finite sample performance of our proposed approach focusing particularly on graph Laplacians. Here, we mainly consider the scenario where the change point locates within the periodic block. The situation when the change point locates at the left boundary of the periodic block is studied in \S S4.2 of the supplementary material. The performance of our method in detecting multiple change points can be found in \S S4.3 of the supplementary material. In \S \ref{subsection::simulation failure} in the supplementary material, we illustrate the limitations when periodic behaviour is disregarded, underscoring the necessity of our approach in such contexts.

\begin{figure}[htbp!]
\centering
\includegraphics[scale=0.42]{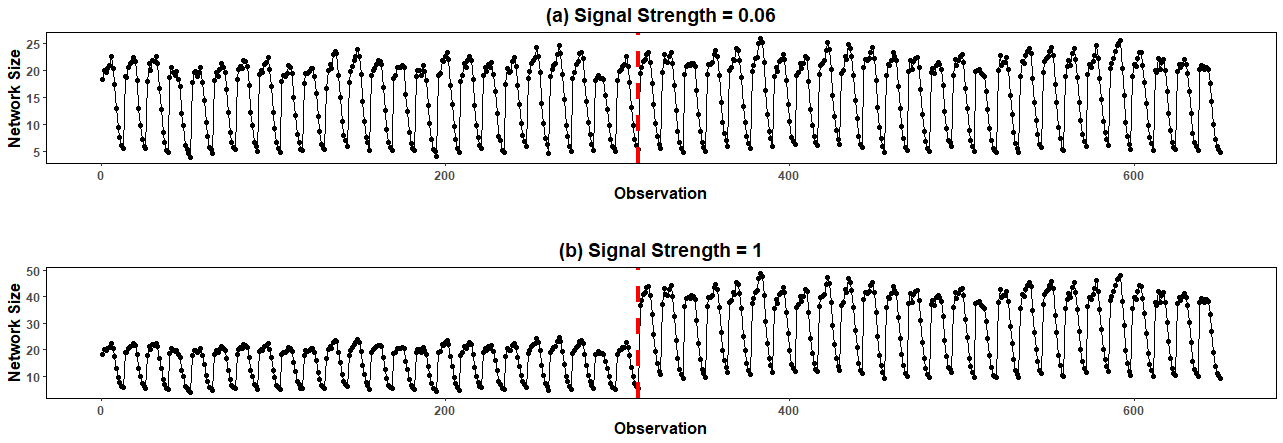}
\caption{The network sizes of the simulated 650 samples with (a) signal strength equal to 0.06 and (b) signal strength equal to 1. The vertical red dashed lines represent the true change point in the middle of the whole sample.  \label{fig::simulate network display}}
\end{figure}

The data generation of periodic networks is given in \S S4.1 in the supplementary material where we consider the period, $M$, of periodic networks is 13. Figure \ref{fig::simulate network display} presents network sizes of simulated 650 networks with the period $M=13$. The clear periodic pattern in network size, evident in Fig. \ref{fig::simulate network display}, confirms that our simulated networks exhibit period $M=13$. Moreover, in the top plot of Fig. \ref{fig::simulate network display}, where the signal strength is $\eta=0.06$, there is no significant difference in network size before and after the true change point, marked by the vertical red dashed line. However, when the signal strength increases to 1, a significant difference in network size becomes apparent, as shown in the bottom plot of Fig. \ref{fig::simulate network display}. Figure \ref{fig::simulate network display} also  underscores the necessity of our approach. Relying solely on network size for change detection would likely result in failure, particularly when changes are subtle. Instead, our method evaluates the random object itself, enabling the detection of changes that would otherwise remain hidden.

The constraint of the scan statistic is set to be $\mathcal{C}_b=[0.1,0.9]$ and the tuning parameter $e_{nn}$ is set to be $\sqrt{K}$, suggested by \cite{chaudhuri2014rates}. The empirical test power is calculated via the proportion of rejections at the nominal level $\alpha=0.05$ across the 200 Monte Carlo simulation runs. For each Monte Carlo run, the permutation scheme in \S \ref{subsect::two step change point} is employed to approximate the critical value of the test with maximum 500 permutations combined with the early stop rule. For the $l$-th Monte Carlo run, let $\hat{\tau}_F^{(l)}$ be the detected change point location. The finite performance of the detected change point locations is assessed by the mean absolute deviation calculated as $\sum_{l=1}^{200} | \hat{\tau}_F^{(l)} - \tau|/200$. 

We now examine the accuracy of our approach when the change point occurs at $Y_{1310}$, the 10th observation of the 100th periodic block. In this scenario, the first 9 observations of the 100th periodic block come from distribution $P_1^M$, while the last 4 observations are drawn from distribution $P_2^M$. 

\begin{figure}[htbp!]
\centering
\includegraphics[scale=0.5]{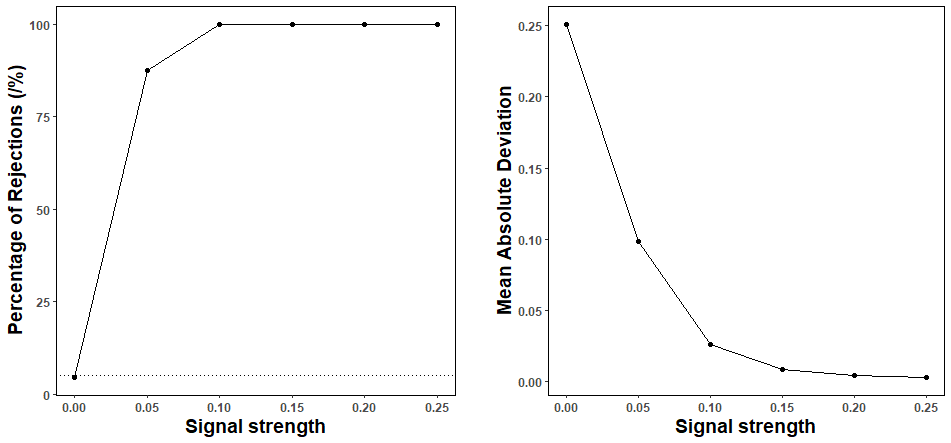}
\caption{The empirical test power (left plot) and the mean absolute deviation of the detected change point locations (right plot) for a sequence of 2600 Laplacian networks with 20 nodes with signal strength varies in $\{0,0.05,\ldots,0.25\}$ and the change point locates within the interior of the periodic block. The horizontal gray dotted line represents the percentage of rejections equal to $5\%$. \label{fig::simulate single cp inside}}
\end{figure}

The left plot of Fig. \ref{fig::simulate single cp inside} displays the empirical power of the proposed test. As expected, the empirical power remains at the significance level $\alpha = 0.05$ under ${\rm H}_0$ when the signal strength is $\eta=0$, and it increases as the signal strength grows. The empirical power reaches its maximum value of 100\% for $\eta \in \{0.1, 0.15, 0.2, 0.25\}$. The right plot of Fig. \ref{fig::simulate single cp inside} illustrates the mean absolute deviation of the estimated change point locations. Our approach demonstrates a decrease in the mean absolute deviation as the signal strength increases. Notably, even though the illustration in Fig. \ref{fig::simulate network display} shows no visible changes in network size at $\eta=0.06$, our method can effectively detect the true change point, supported by Fig. \ref{fig::simulate single cp inside} as the mean absolute deviation is small.
\begin{figure}[!htbp]
\centering
\includegraphics[scale=0.18]{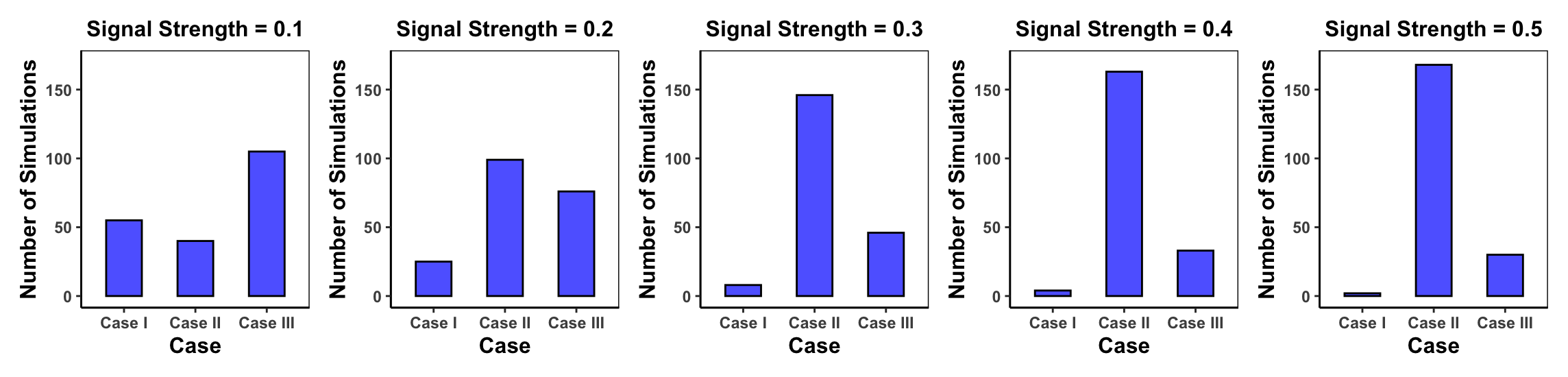}

\caption{Histograms of the simulation results for the number of three cases: (Case I) the estimated change block locates before the true change block, (Case II) the estimated change block is correct, and (Case III) the estimated change block locates after the true change block, considering different signal strengths $\eta\in\{0.1,0.2,0.3,0.4,0.5\}$. The bars present the number of simulations, out of a total of 200, in which one of the three cases happens. \label{fig::hist three cases inside} }
\end{figure}

\begin{figure}[!htbp]
\centering
\includegraphics[scale=0.17]{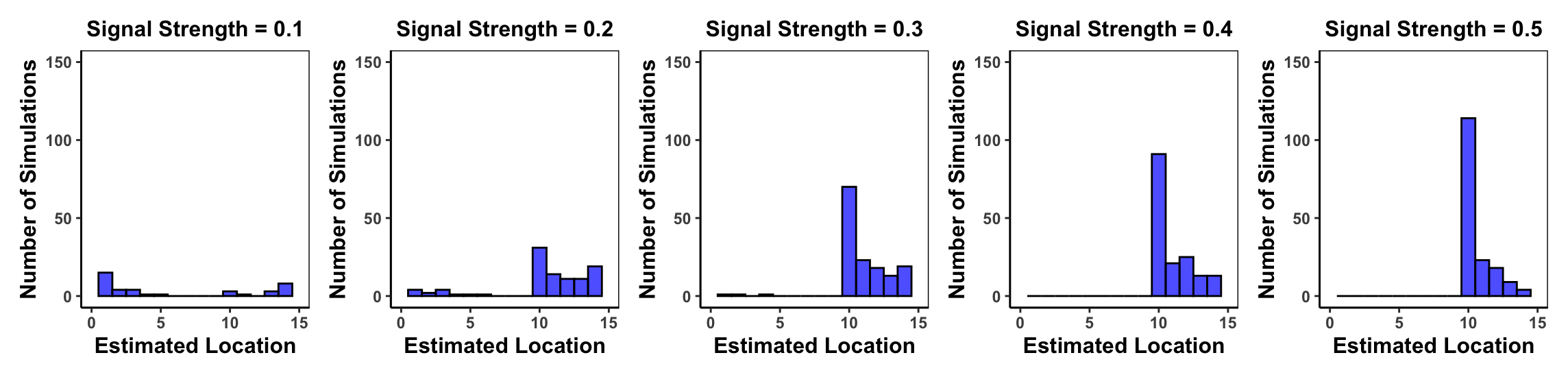}

\caption{ Histograms of the simulation results for the estimated within-block index of the change point when the estimated change block is correct for different signal strengths $\eta\in\{0.1,0.2,0.3,0.4,0.5\}$. The bars present the number of simulations, out of a total of 200, in which a certain value $\hat{\nu}$ is obtained. \label{fig::hist equal inside} }
\end{figure}

To further evaluate the estimated within-block index $\hat{\nu}$, we consider varying signal strengths $\eta\in\{0.1,0.2,0.3,0.4,0.5\}$. Figure \ref{fig::hist three cases inside} shows the occurrence of the three cases: (Case I) the estimated change block locates before the true change block, (Case II) the estimated change block is correct, and (Case III) the estimated change block locates after the true change block. From Fig. \ref{fig::hist three cases inside}, it is not surprising to see that the occurrence of Case II is increasing when the signal strength is larger, which is also supported by Fig. \ref{fig::simulate single cp inside}. Figure \ref{fig::hist equal inside} presents the histogram of the simulation results for $\hat{\nu}$ when the estimated change block is correct. The within-block estimate $\hat{\nu}$ can take values in $\{1,2,\ldots,14\}$ where the value 14 corresponds to the case in which the final estimated change point lies at the left boundary of the block immediately following the initially detected block $\hat{\tau}_b$ in (\ref{formula:: def of change point estimation}). Figure \ref{fig::hist equal inside} demonstrates that $\hat{\nu}$ most frequently equals the true value of 10, with accuracy increasing as the signal strength grows. Similar patterns can be found in the case when the change point locates at the left boundary of the periodic block in \S S4.2 of the supplementary material.

\section{Real Data Analysis}\label{sect::real data}

Historical trip data from the New York City Citi Bike sharing system are publicly available at \url{https://citibikenyc.com/system-data}. This data set records the start and end times of trips, along with start and end locations, at a second-resolution level, encompassing trips between bike stations in New York City. As shown in Fig. \ref{fig::network_display}, there exists clear periodic behaviour in this data.


Our study focuses on trips recorded between October 2019 and May 2020 to assess how public holidays and the COVID-19 pandemic influenced New York City's public transportation networks. By analyzing the hourly dynamics of bike rides across different stations, we aim to uncover patterns within the Citi Bike sharing system and broader transportation trends. We concentrate on the 90 most popular stations, dividing each day into 24 one-hour intervals. During each one-hour interval, a network is constructed with 90 nodes representing the selected stations, where edge weights denote the number of bike trips between station pairs during that interval. This methodology produces a time-varying network spanning 167 days, from October 1, 2019, to May 31, 2020, yielding a total of 4008 observations. 

Each observation corresponds to a 90-dimensional graph Laplacian that encapsulates the network among the 90 stations for a specific one-hour period. The graph Laplacian $L$ for a network with $p$ nodes is obtained as $L = D - A$, where $A$ is the $p \times p$ adjacency matrix, with the $(i,j)$-th entry $a_{ij}$ representing the edge weight between nodes $i$ and $j$, and $D$ denotes the degree matrix, where each diagonal entry is given by $d_{ii} = \sum_{j=1}^p a_{ij}$. 

\begin{figure}[htbp!]
\centering
\includegraphics[scale=0.52]{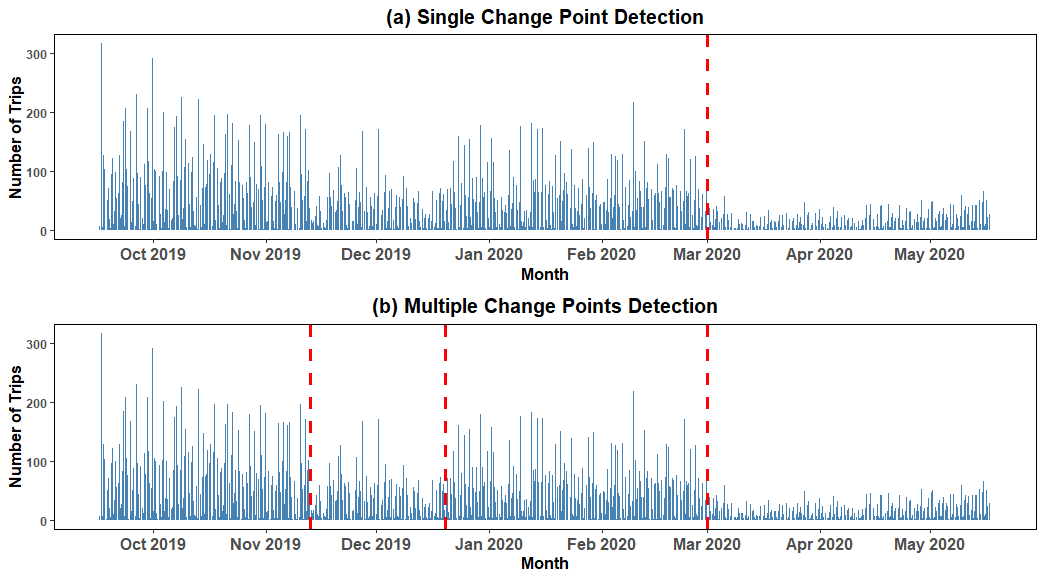}
\caption{The timeline of the number of trips for the whole data with using (a) single change point detection and (b) multiple change point detection. The vertical red dashed lines represent the estimated change points.  \label{fig::real data cp detect}}
\end{figure}

Figure \ref{fig::real data cp detect} illustrates the transportation network sizes of the New York City Citi Bike sharing system, highlighting a noticeable drop in network size during March 2020, coinciding with the COVID-19 outbreak in New York City. By setting the tuning parameter $e_{nn}$ to be $\sqrt{n}$, we first conduct the single change point detection in \S \ref{sect::methodology single}, as shown in Fig. \ref{fig::real data cp detect}(a), identifying a change point on March 16, 2020, at 1 am. On March 15, 2020, the former mayor of New York City, Bill de Blasio announced critical measures in response to the escalating coronavirus crisis, including the closure of the New York City public school system, the largest in the U.S. with 1.1 million students. Following this change point, the number of bike trips plummeted significantly.

We then conduct the multiple change point detection proposed in \S S1 in the supplementary material, with results displayed in Fig. \ref{fig::real data cp detect}(b). In addition to the change point on March 16, 2020, the multiple change point detection algorithm identified additional change points on November 28, 2019, at 1 am, and January 4, 2020, at 6 am.  November 28, 2019 is Thanksgiving, a major public holiday in New York City, and marks the unofficial beginning of the holiday season, characterized by a surge in activities such as Black Friday shopping, followed by Christmas and New Year’s Eve. January 4, 2020, occurs just after New Year’s Day, another public holiday in New York City. It is noteworthy that analyses of Citi Bike’s transportation patterns have shown a general decrease in bike trips from October 2019 to December 2019 due to declining temperatures. However, a sudden drop occurs right after November 27, 2019, likely influenced by the holiday season, with a subsequent increase in trips thereafter. Our proposed test successfully detects this abrupt drop within an overall downward trend.

Each of these change points holds significant relevance: the first on on November 28, 2019 marks the unofficial start of the holiday season after regular working hours, the second on January 4, 2020 signals the end of the holiday season for New Year’s Day, and the last on  March 16, 2020 corresponds to the beginning of New York State’s shutdown measures aimed at curbing the spread of COVID-19. Notably, while the second change point falls within the corresponding periodic block, the first and third lie at the left boundary of their respective blocks. Periodic patterns of weekday and holiday observations typically differ, explaining why the first detected change point occurs at the beginning of the periodic block, as it corresponds to a location where that the periodic behaviour changes. Similarly, the third change point lies at the left boundary due to the distinct changes in periodic behaviour following the implementation of the shutdown measures.

\section{Discussion}

The proposed framework permits a flexible choice of change point detector for random objects. For example, if the main aim to to test the change in mean, rather than the change in distribution, we could consider using a Fr\'{e}chet-mean-based detector; see, e.g., \cite{dubey2020frechet} and \cite{jiang2024two}. However, the irregular situation of the change block after the rearrangement poses additional theoretical challenges as the data segment after the change point is not identically distributed, which is a commonly used assumption in the literature. Therefore,  detailed theoretical development of any suitable change point detector in our proposed framework is needed, which can follow the proof strategy in the supplementary material.

This paper identifies the existence of periodic patterns in random objects and demonstrates that ignoring such periodicity can negatively impact change point detection. Periodicity is a fundamental characteristic of time-varying data; however, periodic patterns in time-varying random objects remain relatively underexplored. In this paper, we assume that the period is known, as is the case for the real data application we consider, where the period is clearly defined. Nevertheless, in many practical scenarios, the period may be unknown. Addressing change point detection under unknown periodicity is an important direction for future research.

\section*{Acknowledgement}
The work of Andrew T. A. Wood was supported by Australian Research Council grant DP220102232. The work of Tao Zou was supported by the assistance of computational resources provided by the Australian Government through the National Computational Infrastructure (NCI) under the ANU Startup Allocation Scheme.

\section*{Supplementary material}
The supplementary material consists of the multiple change point detection procedure for periodic random objects, theoretical properties of the proposed multiple change point detection procedure, several technical lemmas, the proofs of the main results, additional simulation settings and results for single change point detection and multiple change point detection, and a comparison between the metric distribution function and distance profiles. 

\appendix

\appendixone
\section*{Appendix}
We first define directionally $(\epsilon,\varsigma,L)$-limited metric, which is a geometrical condition on the metric.  Then we state the assumptions used to derive our theoretical results.

\begin{definition}[\citealt{federer2014geometric}]
A metric $d$ is called directionally $(\epsilon,\varsigma,L)$-limited at the subset $\Omega_{\rm sub}$ of $\Omega$, if the following condition holds: for $\epsilon>0,0<\varsigma\leq 1/3$ and a positive integer $L$, if for each $\omega\in\Omega_{\rm sub}$ and $D\subseteq \Omega_{\rm sub}\cap \bar{B}(\omega,\epsilon)$ such that $d(x,a)\geq \varsigma d(\omega,a)$ whenever $a,b\in D$, $a\neq b$, $x\in\Omega$ with
$
d(\omega,x) = d(\omega,a)$, $d(x,b)+d(x,\omega)=d(\omega,b),
$
then the cardinality of $D$ is no larger than L.
\end{definition}

\begin{assumption}\label{assump 1}
Suppose that $(\Omega,d)$ is a a complete separable metric space and the metric $d$ is directionally $(\epsilon,\varsigma,L)$-limited at the support of the probability measure $P$. If, in addition, $\Omega$ is an infinite dimension space, then for any $\varepsilon$, suppose that there exists $\Omega_l\subseteq \Omega$ such that $P(\Omega_l)\leq 1-\varepsilon$ and the metric $d$ is directionally $(\epsilon(\Omega_l),\varsigma(\Omega_l),L(\Omega_l))$-limited at $\Omega_l$.
\end{assumption}

\begin{assumption}\label{assump 2}
Denote $N(\delta,\Omega,d)$ as the covering number of the metric space $(\Omega,d)$ with balls of radius $\delta$. Then $\delta \log N(\delta,\Omega,d)\to 0$ as $\delta\to 0$.
\end{assumption}

\begin{assumption}\label{assump 3}
Let $d^M$ be the metric of the product space $\Omega^M$ where $d^M(\bm{w},\bm{v})=\sqrt{\sum_{m=1}^M d^2(u_m,v_m)}$ with $\bm{w},\bm{v}\in\Omega^M$, $\bm{w}=(u_1,\ldots,u_M)$ and $\bm{v}=(v_1,\ldots,v_M)$. Assume that for  $\bm{Z}\sim P_1^M$ and $\bm{Z}^\prime\sim P_2^M$, the cumulative distribution functions of $d^M(\bm{w},\bm{Z})$ and $d^M(\bm{w},\bm{Z}^\prime)$ have probability density functions $f_{\bm{w}}^{(1)}(t)$ and $f_{\bm{w}}^{(2)}(t)$ for all $\bm{w}\in\Omega^M$, and satisfy $\sup_{\bm{w}\in\Omega^M}\sup_{t\in\mathbb{R}}\left| f_{\bm{w}}^{(1)}(t)\right|\leq C_{f,1}<\infty$ and $\sup_{\bm{w}\in\Omega^M}\sup_{t\in\mathbb{R}}\left| f_{\bm{w}}^{(2)}(t)\right|\leq C_{f,2}<\infty$ for some constants $C_{f,1}$ and $C_{f,2}$.
\end{assumption}

\begin{assumption}\label{assump 4}
There exists a constant $C_w>0$ such that $\sup_{\bm{x},\bm{y}\in\Omega^M}|W(\bm{x},\bm{y})|\leq C_w$.
\end{assumption}

Assumption \ref{assump 1} involves a geometrical condition on the metric, called directionally $(\epsilon,\varsigma,L)$-limited, that was introduced by \cite{federer2014geometric}. Assumption \ref{assump 1} is used to establish the one-to-one correspondence of the metric distribution function (\citealt{wang2023nonparametric}). Assumption \ref{assump 2} controls the complexity of the metric space $(\Omega,d)$ and is a standard condition in empirical process theory; see van der Vaart and Wellner (2013).  Assumption \ref{assump 3} imposes regularity conditions on the metric distribution function under the distributions $P_1^M$ and $P_2^M$. Assumptions \ref{assump 1} and \ref{assump 2} are applicable to many metric spaces including spaces of smooth functions, Riemannian manifold spaces, shape spaces and the space of phylogenetic trees (\citealt{dubey2024metric,dubey2023change,wang2023nonparametric}). Assumptions \ref{assump 2} and \ref{assump 3} are used to establish the Glivenko-Cantelli and Donsker properties, as defined in van der Vaart and Wellner (2013), of the empirical metric distribution function (\citealt{wang2023nonparametric}). Assumption \ref{assump 4} is a regularity condition on the data-adaptive weight function $W$, to ensure well-behaved asymptotic limit distributions.

\bibliographystyle{biometrika}

\begin{thebibliography}{7}
\expandafter\ifx\csname natexlab\endcsname\relax\def\natexlab#1{#1}\fi

\bibitem[Aminikhanghahi and Cook, 2017]{aminikhanghahi2017survey}
Aminikhanghahi, S. and Cook, D.~J. (2017).
\newblock A survey of methods for time series change point detection.
\newblock {\em Knowledge and Information Systems}, 51(2):339--367.

\bibitem[Andreou and Ghysels, 2002]{andreou2002detecting}
Andreou, E. and Ghysels, E. (2002).
\newblock Detecting multiple breaks in financial market volatility dynamics.
\newblock {\em Journal of Applied Econometrics}, 17(5):579--600.



\bibitem[Bai and Perron, 2003]{bai2003computation}
Bai, J. and Perron, P. (2003).
\newblock Computation and analysis of multiple structural change models.
\newblock {\em Journal of Applied Econometrics}, 18(1):1--22.



\bibitem[Basseville et~al., 1993]{basseville1993detection}
Basseville, M., Nikiforov, I.~V., et~al. (1993).
\newblock {\em Detection of Abrupt Changes: Theory and Application}, volume
  104.
\newblock prentice Hall Englewood Cliffs.

\bibitem[Beaulieu and Killick, 2018]{beaulieu2018distinguishing}
Beaulieu, C. and Killick, R. (2018).
\newblock Distinguishing trends and shifts from memory in climate data.
\newblock {\em Journal of Climate}, 31(23):9519--9543.



\bibitem[Chaudhuri and Dasgupta, 2014]{chaudhuri2014rates}
Chaudhuri, K. and Dasgupta, S. (2014).
\newblock Rates of convergence for nearest neighbor classification.
\newblock {\em Advances in Neural Information Processing Systems}, 27.

\bibitem[Chen and Friedman, 2017]{chen2017new}
Chen, H. and Friedman, J. H. (2017).
\newblock A new graph-based two-sample test for multivariate and object data.
\newblock {\em Journal of the American Statistical Association}, 112(517):397--409.

\bibitem[Chen and Zhang, 2015]{chen2015graph}
Chen, H. and Zhang, N. (2015).
\newblock Graph-based change-point detection.
\newblock {\em The Annals of Statistics}, 43(1):139--176.

\bibitem[Chung and Romano, 2013]{chung2013exact}
Chung, E. and Romano, J.~P. (2013).
\newblock Exact and asymptotically robust permutation tests.
\newblock {\em The Annals of Statistics}, 41(2):484--507.

\bibitem[Chung and Romano, 2016]{chung2016asymptotically}
Chung, E. and Romano, J.~P. (2016).
\newblock Asymptotically valid and exact permutation tests based on two-sample
  $U$-statistics.
\newblock {\em Journal of Statistical Planning and Inference}, 168:97--105.

\bibitem[de~la Pena and Montgomery-Smith, 1995]{de1995decoupling}
de~la Pena, V.~H. and Montgomery-Smith, S.~J. (1995).
\newblock Decoupling inequalities for the tail probabilities of multivariate $U$-statistics.
\newblock {\em The Annals of Probability}, pages 806--816.

\bibitem[Dubey et~al., 2024]{dubey2024metric}
Dubey, P., Chen, Y., and M{\"u}ller, H.-G. (2024).
\newblock Metric statistics: exploration and inference for random objects with distance profiles.
\newblock {\em The Annals of Statistics}, 52(2):757--792.

\bibitem[Dubey and M{\"u}ller, 2020]{dubey2020frechet}
Dubey, P. and M{\"u}ller, H.-G. (2020).
\newblock Fr{\'e}chet change-point detection.
\newblock {\em The Annals of Statistics}, 48(6):3312--3335.


\bibitem[Dubey and Zheng, 2023]{dubey2023change}
Dubey, P. and Zheng, M. (2023).
\newblock Change point detection for random objects using distance profiles.
\newblock {\em arXiv preprint arXiv:2311.16025}.


\bibitem[Dümbgen, 1991]{dumbgen1991asymptotic}
D{\"u}mbgen, L. (1991).
\newblock The asymptotic behavior of some nonparametric change-point estimators.
\newblock {\em The Annals of Statistics}, 19(3), 1471--1495.


\bibitem[Eichelsbacher, 2001]{eichelsbacher2001moderate}
Eichelsbacher, P. (2001).
\newblock Moderate deviations for functional U-processes.
\newblock In {\em Annales de l'Institut Henri Poincare (B) Probability and
  Statistics}, volume~37, pages 245--273. Elsevier.

\bibitem[Erdman and Emerson, 2008]{erdman2008fast}
Erdman, C. and Emerson, J.~W. (2008).
\newblock A fast Bayesian change point analysis for the segmentation of
  microarray data.
\newblock {\em Bioinformatics}, 24(19):2143--2148.

\bibitem[Erlemann et al.(2022)]{erlemann2022cramer}
Erlemann, R., Lockhart, R., and Yao, R. (2022).
Cram{\'e}r-von Mises tests for change points.
\textit{Scandinavian Journal of Statistics}, 49(2), 802--830.



\bibitem[Fearnhead and Rigaill, 2019]{fearnhead2019changepoint}
Fearnhead, P. and Rigaill, G. (2019).
\newblock Changepoint detection in the presence of outliers.
\newblock {\em Journal of the American Statistical Association},
  114(525):169--183.

\bibitem[Federer, 2014]{federer2014geometric}
Federer, H. (2014).
\newblock {\em Geometric Measure Theory}.
\newblock Springer.

\bibitem[Fryzlewicz, 2014]{fryzlewicz2014wild}
Fryzlewicz, P. (2014).
\newblock Wild binary segmentation for multiple change-point detection.
\newblock {\em Annals of Statistics}, 42(6):2243--2281.



\bibitem[Guo and Modarres, 2020]{guo2020nonparametric}
Guo, L. and Modarres, R. (2020).
\newblock Nonparametric change point detection for periodic time series.
\newblock {\em Canadian Journal of Statistics}, 48(3):518--534.

\bibitem[Hapfelmeier et~al., 2023]{hapfelmeier2023efficient}
Hapfelmeier, A., Hornung, R., and Haller, B. (2023).
\newblock Efficient permutation testing of variable importance measures by the
  example of random forests.
\newblock {\em Computational Statistics \& Data Analysis}, 181:107689.





\bibitem[Hawkins et~al., 2003]{hawkins2003changepoint}
Hawkins, D.~M., Qiu, P., and Kang, C.~W. (2003).
\newblock The changepoint model for statistical process control.
\newblock {\em Journal of Quality Technology}, 35(4):355--366.

\bibitem[Holmes, 2003]{holmes2003statistics}
Holmes, S. (2003).
\newblock Statistics for phylogenetic trees.
\newblock {\em Theoretical Population Biology}, 63(1):17--32.

\bibitem[Jiang et~al., 2023]{jiang2023time}
Jiang, F., Zhao, Z., and Shao, X. (2023).
\newblock Time series analysis of Covid-19 infection curve: a change-point perspective.
\newblock {\em Journal of Econometrics}, 232(1):1--17.

\bibitem[Jiang et al., 2024]{jiang2024two}
Jiang, F., Zhu, C., and Shao, X. (2024).
\newblock Two-sample and change-point inference for non-Euclidean valued time series.
\newblock {\em Electronic Journal of Statistics}, 18(1), 848--894.

\bibitem[Kanrar et al., 2024]{kanrar2024model}
Kanrar, R., Jiang, F., and Cai, Z. (2024).
\newblock Model-free change-point detection using modern classifiers.
\newblock {\em arXiv preprint arXiv:2404.06995}.


\bibitem[Kojadinovic and Verdier, 2021]{kojadinovic2021nonparametric}
Kojadinovic, I. and Verdier, G. (2021).
\newblock Nonparametric sequential change-point detection for multivariate time
  series based on empirical distribution functions.
\newblock {\em Electronic Journal of Statistics}, 15(1):773--829.


\bibitem[Kolar et~al., 2010]{kolar2010estimating}
Kolar, M., Song, L., Ahmed, A., and Xing, E.~P. (2010).
\newblock Estimating time-varying networks.
\newblock {\em The Annals of Applied Statistics}, 94--123.

\bibitem[Kosorok, 2008]{kosorok2008introduction}
Kosorok, M.~R. (2008).
\newblock {\em Introduction to Empirical Processes and Semiparametric
  Inference.}
\newblock Springer.

\bibitem[Kossinets and Watts, 2006]{kossinets2006empirical}
Kossinets, G. and Watts, D.~J. (2006).
\newblock Empirical analysis of an evolving social network.
\newblock {\em Science}, 311(5757):88--90.

\bibitem[Kov{\'a}cs et~al., 2023]{kovacs2023seeded}
Kov{\'a}cs, S., B{\"u}hlmann, P., Li, H., and Munk, A. (2023).
\newblock Seeded binary segmentation: a general methodology for fast and
  optimal changepoint detection.
\newblock {\em Biometrika}, 110(1):249--256.

\bibitem[Lai, 1995]{lai1995sequential}
Lai, T.~L. (1995).
\newblock Sequential changepoint detection in quality control and dynamical
  systems.
\newblock {\em Journal of the Royal Statistical Society: Series B
  (Methodological)}, 57(4):613--644.


\bibitem[Lee, 2019]{lee2019u}
Lee, A.~J. (2019).
\newblock {\em $U$-statistics: Theory and Practice}.
\newblock Routledge.



\bibitem[Marron and Dryden, 2021]{marron2021object}
Marron, J.~S. and Dryden, I.~L. (2021).
\newblock {\em Object Oriented Data Analysis}.
\newblock Chapman and Hall/CRC.

\bibitem[Matteson and James, 2014]{matteson2014nonparametric}
Matteson, D.~S. and James, N.~A. (2014).
\newblock A nonparametric approach for multiple change point analysis of multivariate data.
\newblock {\em Journal of the American Statistical Association},
  109(505):334--345.



\bibitem[Olshen et~al., 2004]{olshen2004circular}
Olshen, A.~B., Venkatraman, E.~S., Lucito, R., and Wigler, M. (2004).
\newblock Circular binary segmentation for the analysis of array-based DNA copy number data.
\newblock {\em Biostatistics}, 5(4):557--572.


\bibitem[Pan et al.(2020)]{pan2020ball}
Pan, W., Wang, X., Zhang, H., Zhu, H., and Zhu, J. (2020).
Ball covariance: a generic measure of dependence in Banach space.
\textit{Journal of the American Statistical Association}, 115(529), 307--317.

\bibitem[Reeves et~al., 2007]{reeves2007review}
Reeves, J., Chen, J., Wang, X.~L., Lund, R., and Lu, Q.~Q. (2007).
\newblock A review and comparison of changepoint detection techniques for climate data.
\newblock {\em Journal of Applied Meteorology and Climatology}, 46(6):900--915.

\bibitem[Rodionov, 2004]{rodionov2004sequential}
Rodionov, S.~N. (2004).
\newblock A sequential algorithm for testing climate regime shifts.
\newblock {\em Geophysical Research Letters}, 31(9).

\bibitem[Scealy and Wood, 2023]{scealy2023score}
Scealy, J.~L. and Wood, A.~T.~A. (2023).
\newblock Score matching for compositional distributions.
\newblock {\em Journal of the American Statistical Association},
  118(543):1811--1823.

\bibitem[Shewhart and Deming, 1986]{shewhart1986statistical}
Shewhart, W.~A. and Deming, W.~E. (1986).
\newblock {\em Statistical Method from the Viewpoint of Quality Control}.
\newblock Courier Corporation.


\bibitem[Truong et~al., 2020]{truong2020selective}
Truong, C., Oudre, L., and Vayatis, N. (2020).
\newblock Selective review of offline change point detection methods.
\newblock {\em Signal Processing}, 167:107299.



\bibitem[van der Vaart, 1998]{van2000asymptotic}
van der Vaart, A. (1998).
\newblock \textit{Asymptotic Statistics}.
\newblock Cambridge University Press.

\bibitem[van~der Vaart and Wellner, 2013]{wellner2013weak}
van~der Vaart, A. and Wellner, J. (2013).
\newblock {\em Weak Convergence and Empirical Processes: with Applications to
  Statistics}.
\newblock Springer Science \& Business Media.





\bibitem[Wang et al., 2022]{wang2022robust}
Wang, S., Huang, T., You, J., and Cheng, M.-Y. (2022).
\newblock Robust inference for nonstationary time series with possibly multiple changing periodic structures.
\newblock {\em Journal of Business \& Economic Statistics}, 40(4):1718--1731.

\bibitem[Wang et~al., 2024]{wang2023nonparametric}
Wang, X., Zhu, J., Pan, W., Zhu, J., and Zhang, H. (2024).
\newblock Nonparametric statistical inference via metric distribution function in metric spaces.
\newblock {\em Journal of the American Statistical Association},
  119(548), 2772-2784.

\bibitem[Wang et~al., 2017]{wang2017fast}
Wang, Y., Chakrabarti, A., Sivakoff, D., and Parthasarathy, S. (2017).
\newblock Fast change point detection on dynamic social networks.
\newblock {\em arXiv preprint arXiv:1705.07325}.


\bibitem[Werenski et al., 2024]{werenski2024rank}
Werenski, M., Masud, S. B., Murphy, J. M., and Aeron, S. (2024).
\newblock On rank energy statistics via optimal transport: Continuity, convergence, and change point detection.
\newblock {\em IEEE Transactions on Information Theory}.

\bibitem[Worsley et~al., 2002]{worsley2002general}
Worsley, K.~J., Liao, C.~H., Aston, J., Petre, V., Duncan, G., Morales, F., and
  Evans, A.~C. (2002).
\newblock A general statistical analysis for fMRI data.
\newblock {\em Neuroimage}, 15(1):1--15.


\bibitem[Zhang et al., 2017]{zhang2017pruning}
Zhang, W., James, N. A., and Matteson, D. S. (2017).
\newblock Pruning and nonparametric multiple change point detection.
\newblock In {\em Proceedings of the 2017 IEEE International Conference on Data Mining Workshops (ICDMW)}, pages 288--295. IEEE.

\bibitem[Zhang et al., 2025]{zhang2025change}
Zhang, Y., Zhu, C., and Shao, X. (2025).
\newblock Change-point detection for object-valued time series.
\newblock {\em Journal of Business \& Economic Statistics}, (just-accepted):1--23. 



\end{thebibliography}

\end{document}